\documentstyle[epsfig]{disk}
\begin{document}

\title{%
Disc and secondary irradiation 
in dwarf and X-ray novae}

\author{Jean-Pierre Lasota \\
{\it DARC, Observatoire de Paris,\\
Section de Meudon, 92190 Meudon, France\\ 
jpl@orge.obspm.fr}\\
}

\maketitle

\section*{Abstract}
I present a short review of irradiation processes in close binary systems.

\section{Introduction}

Thermal-viscous instability is widely accepted as the origin of Dwarf Nova and
Soft X-ray Transient (X-ray Nova) outbursts. In the `standard' model, one assumes a constant 
mass-transfer rate, no effects of irradiation are taken into account and the accretion disc is
supposed to extend down to the surface of the accreting body (or the last stable orbit in the
case of accreting black holes and compact neutron stars). This version of the model, however,
explains neither all the observed outburst types nor all the observed outburst
properties. Generalizations of the `standard' model which take into account some or all of these
neglected, but obviously important, effects have been proposed and studied in various 
investigations. I will shortly discuss some recent results concerning irradiation
(Hameury, Lasota \&  Hur\'e 1997; Dubus et al. 1999; hereafter DLHC;
King (1997); King \& Ritter 1998). 

Since in the literature of the subject most articles use incorrect equations to describe
irradiated  accretion discs I begin
by repeating the simple derivation of the basic equations presented in DLHC.

\section{A simple introduction to the vertical structure of irradiated discs}

In accretion discs the vertical energy conservation equation has the form:
\begin{equation}
{dF \over dz} = Q_{\rm vis}(R,z)
\label{energy1}
\end{equation}
where $F$ is the vertical (in the $z$ direction) radiative flux and $Q_{\rm
vis}(R,z)$ is the viscous heating rate per unit volume. Eq. (\ref{energy1})
states that an accretion disc is not in radiative equilibrium, contrary to a
stellar atmosphere. 
Using the ``$\alpha$ - viscosity'' prescription (Shakura \& Sunyaev 1973) $\nu =
(2/3)\alpha c_{\rm s}^2/\Omega_{\rm K}$, where $\alpha$ is the viscosity
parameter ($\leq 1$), $\Omega_{\rm K}$ is the Keplerian angular frequency and
$ c_{\rm s}=\sqrt{P/\rho}$ is the sound speed, $\rho$ the density, and $P$
the pressure one can write
\begin{equation}
Q_{\rm vis}(R,z)= (3/2) \alpha \Omega_{\rm K} P
\label{voldiss}
\end{equation}
Viscous heating of this form has important implications for the structure of
optically thin layers of accretion discs and may lead to creation of coronae
and winds (Shaviv \& Wehrse 1986; 1991). Here, however, we are interested in
the effects of irradiation on the inner structure of an optically thick disc,
so our results should depend on the precise form of the viscous heating.
We neglect, however, the possible presence of
an X-ray irradiation generated corona and wind, described by Idan \& Shaviv
(1996).
\begin{figure}[t]
\centering\epsfig{figure=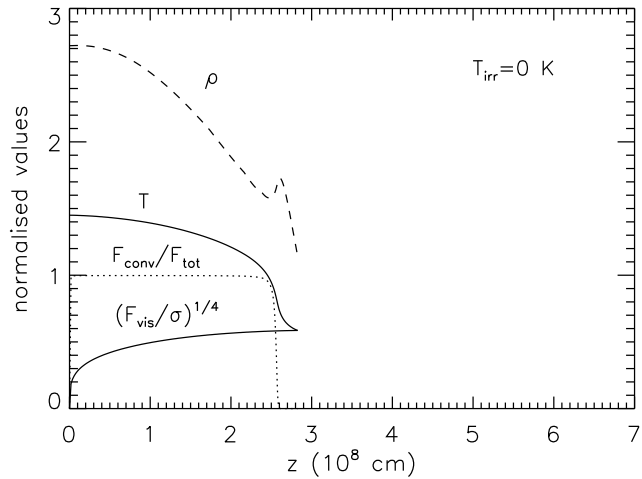,height=7cm,
}      
   \caption{
Vertical structure of an accretion disc around a $M=10 M_{\odot}$
compact object at $r =3\times 10^{10}$ cm. $\alpha\approx 0.1$, $\dot M
\approx 10^{16}$ g s$^{-1}$ (i.e. $T_{\rm eff}\approx 5700$ K) and $T_{\rm
irr}=0$. Both temperatures are normalized by $10^4$
K. Since $T_{\rm irr}=0$, the surface temperature $T(\tau_{\rm s})= T_{\rm
eff}$. The dashed line is the density in units of $ \times 10^{-7}$ g
cm$^{-3}$, the dotted line the ratio of the convective to the
total fluxes. Note that for these parameters, the section of the disc lays on
the lower stable branch. (From DLHC)}
\end{figure}
\begin{figure}[t]
\centering
\epsfig{figure=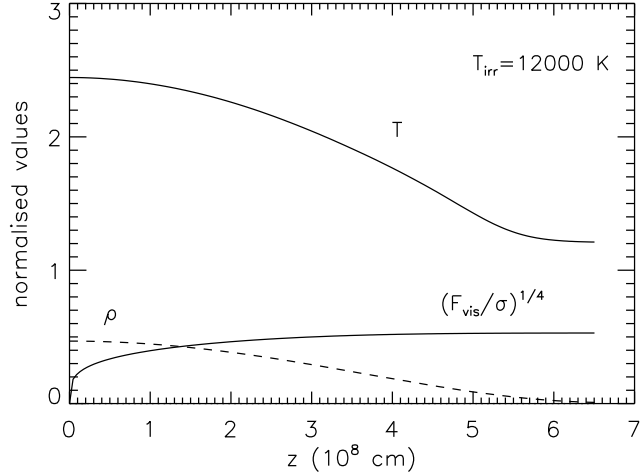,height=7cm,
}      
   \caption{Same as Fig. 1. but for $T_{\rm irr}=12000$ K. (From DLHC) } 
\end{figure}
When integrated over $z$, the rhs of Eq. (\ref{energy1}) is equal to viscous
dissipation per unit surface:
\begin{equation}
F_{\rm vis}={3 \over 2} \alpha \Omega_{\rm K} \int_0^{+\infty} P dz ,
\label{fvis}
\end{equation}
which is close, but not exactly equal, to the surface heating term $(9/8) \nu
\Sigma \Omega_{\rm K}^2$ generally used in the literature. The difference
between the two expressions may be important in numerical calculation
(Hameury et al. 1998) but in the present context is of no importance.
One can rewrite Eq. (\ref{energy1}) as
\begin{equation}
{dF\over d\tau} = - f(\tau ){F_{\rm vis} \over \tau_{\rm T}}
\label{energy2}
\end{equation}
where we introduced a new variable, the optical depth $d\tau=-\kappa_{\rm R}
\rho dz$, $\kappa_{\rm R}$ being the Rosseland mean opacity and $\tau_{\rm T}
= \int_0^{+\infty} \kappa_{\rm R} \rho dz$ is the total optical depth.
As shown in DLHC, putting $f(\tau)=1$ is a good approximation.

At the disc midplane, by symmetry, the flux must vanish: $F(\tau_{\rm T})=0$,
whereas at the surface, ($\tau=0$)
\begin{equation}
F(0) \equiv \sigma T^4_{\rm eff}= F_{\rm vis}
\label{fsurface}
\end{equation}
Equation (\ref{fsurface}) states that the total flux at the surface is equal
to the energy dissipated by viscosity (per unit time and unit surface). The
solution of Eq. (\ref{energy2}) is thus
\begin{equation}
F(\tau) = F_{\rm vis} \left(1 - {\tau\over \tau_{\rm T}}\right)
\label{flux0}
\end{equation}
where $ \tau_{\rm tot}$ is the total
optical depth. 

To obtain the temperature stratification one has to solve the transfer
equation. Here we use the diffusion approximation
\begin{equation}
F(\tau) = {4 \over 3} {\sigma dT^4 \over d\tau} ,
\label{diff}
\end{equation}
appropriate for the optically thick discs we are dealing with. The
integration of Eq. (\ref{diff}) is straightforward and gives :
\begin{equation}
T^4(\tau) - T^4(0) = {3\over 4} \tau \left(1 - {\tau \over 2\tau_{\rm T}}
			\right) T^4_{\rm eff}
\label{t1}
\end{equation}

The upper (surface) boundary condition is:
\begin{equation}
T^4(0) = {1 \over 2} T^4_{\rm eff} + T^4_{\rm irr}
\label{bcond2}
\end{equation}
where $T^4_{\rm irr}$ is the irradiation temperature, which depends on $R$,
the albedo, the height at which the energy is deposited and on the shape of
the disc (see Eq. \ref{tirr}). In Eq. (\ref{bcond2}) $T(0)$ corresponds to
the {\sl emergent} flux and, as mentioned above, $T_{\rm eff}$ corresponds to
the {\sl total} flux, hence the factor 1/2 in front
of $T^4_{\rm eff}$. The temperature stratification
is thus :
\begin{equation}
T^4(\tau) ={3\over 4} T^4_{\rm eff}\left[ \tau \left(1 - {\tau \over 2\tau_{\rm T}}\right)
		+ {2 \over 3}\right]  + T^4_{\rm irr}.
\label{t2}
\end{equation}
For $\tau_{\rm T} \gg 1$, the temperature
at the disc midplane is
\begin{equation}
T^4_{\rm c} \equiv T^4(\tau_{\rm T}) =
		 {3 \over 8} \tau_{\rm tot} T_{\rm eff}^4 + T^4_{\rm irr}
\label{diff2}
\end{equation}
It is clear, therefore, that for the disc inner structure to be dominated by
irradiation and the disc to be isothermal one must have
\begin{equation}
{F_{\rm irr}\over \tau_{\rm tot}} \equiv {\sigma T^4_{\rm irr} \over
\tau_{\rm tot}} \gg F_{\rm vis}
\label{c1}
\end{equation}
and not just $F_{\rm irr} \gg F_{\rm vis}$ as is usually assumed. The
difference between the two criteria is important in low-mass X-ra binaries since, for
parameters of interest, $\tau_{\rm tot} \sim 10^2 - 10^3$ in the outer disc
regions. 

The effect of disc irradiation is illustrated on Figs. 1 \& 2 (DLHC). 
Fig. 1 shows the
vertical structure of a ring which is part of an unilluminated accretion disc.
This ring is on the lower, cool branch of the S-curve. The energy transport is
dominated by convection. The surface temperature is equal to the effective
temperature given by viscous dissipation (see Eq. \ref{fsurface}). The vertical
structure of an irradiated disc is shown on Fig. 2. Although the irradiating
flux is 20 times larger than the viscous flux the disc is not isothermal.
Since in the non-irradiated disc $T_{\rm c} \approx 14500$ K and the surface
temperature is $T_{\rm s} \approx 5700$ the optical depth is $\sim 100$ and, obviously
an irradiation temperature higher than 14500 K is required to
make the disc isothermal. 
Note that irradiation suppressed convection and the disc now is purely radiative.

\begin{figure}[t]
\centering\epsfig{figure=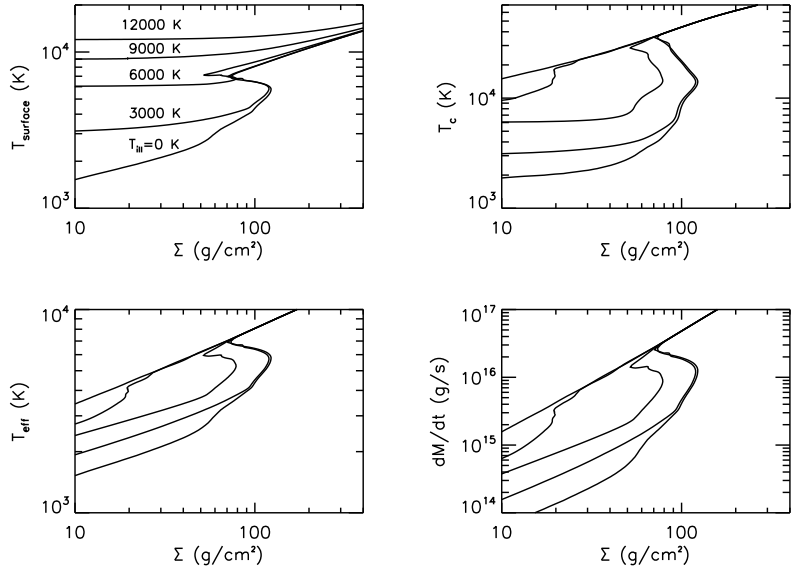,height=8cm,
}  
\caption{$\Sigma$ -- $T_{\rm surface}$, $\Sigma$ -- $T_{\rm c}$, $\Sigma$ --
$T_{\rm eff}$ and $\Sigma$ -- $\dot M$ curves for $r =3\cdot 10^{10}$ cm,
$M=10M_{\odot}$, $\alpha$ = 0.1, and $T_{\rm irr}=
[0,3,6,9,12]\times 10^3$ K. (From DLHC)}
\end{figure}

\begin{figure}[t]
\centering\epsfig{figure=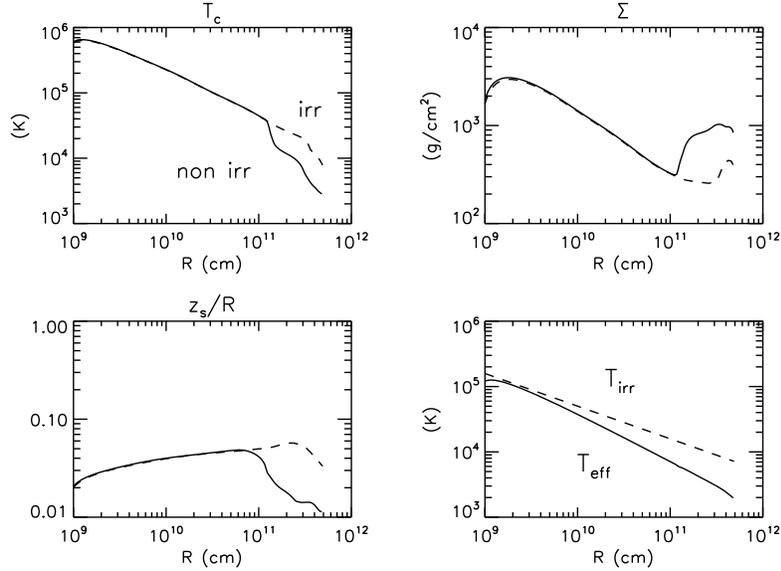,height=8cm,
}
\caption{Radial profiles of the midplane temperature, surface density, and
photospheric height to radius ratio for an un-irradiated (continuous line)
and irradiated disc (dashed line) around a 10 M$_\odot$ compact object. In
the temperature diagram the continuous line represents the effective
temperature. The accretion rate is $\dot M= 10^{18}$ g s$^{-1}$,
$T_{\rm irr}$ is taken from Eq. (15) with ${\cal C} = 5 \times 10^{-4}$.
Regions beyond the radius at which a break in the $T_{\rm c}$ or $\Sigma$
curve appears are unstable. (From DLHC)}
\end{figure}

\section{Can the outer disc `see' a point source located at the midplane ?}

The answer is `no' (see e.g. DLHC). The reason is that contrary to the frequently
made assumption a `standard' accretion disc is convex rather than concave.
Images showing flaring discs have nothing to do with reality, at least with
reality described by the standard model of planar discs. 

For a point source, the irradiation temperature can be written as
\begin{equation}
T^4_{\rm irr} = {\eta \dot Mc^2 (1 - \varepsilon) \over 4 \pi \sigma
		R^2}{H_{\rm irr} \over R} \left({d\ln H_{\rm irr}
		\over d\ln R} - 1\right)
\label{tirr}
\end{equation}
where $\eta$ is the efficiency of converting accretion power into X-rays,
$\varepsilon$ is the X-ray albedo and $H_{\rm irr}$ is the local height at
which irradiation energy is deposited, or the height of the disc ``as seen by
X-rays". We use here $H_{\rm irr}$ and {\sl not} $H$, the local pressure
scale-height, as is usually written in the literature because, in general,
$H_{\rm irr}\neq H$. 

Eq. (\ref{tirr}) is usually used (at least in the recent, abundant, publications
on the subject) with `typical' values of $ \varepsilon > 0.9$, $H/R=0.2$
(no difference is seen between $H_{\rm irr}$ and $H$; the fact that there is no
reason for the photosphere to be at one (isothermal) scale-height seems to be
largely ignored). These values are supposed to be given by `observations'. 
However, when one reads the articles 
quoted to support this assertion (Vrtilek et al. 1990; de Jong et al. 1996) 
one finds nothing of the kind there. These
papers {\sl assume} that an irradiated disc is isothermal. de Jong et al. (1996)
fit light-curves with the isothermal model of Vrtilek et al. (1990).
Moreover, since de Jong et al. (1996) model lightcurves of neutron-star binaries
the $H/R=0.2$ cannot be applied to black-hole binaries 
(especially if $H$ were the pressure
scale-height): since black-holes are more massive than neutron stars, in their
case $H/R$ should be smaller at a given radius (let me add for the benefit of
some readers: this is because gravity is then stronger), 
as seen in Fig. 6 \& 7 of DLHC. In any case, for $H/R=0.2$ the vertical 
hydrostatic equilibrium equation would imply high temperatures at the disc's outer
rim:
\begin{equation}
T \sim 8 \times 10^7 \left({M \over M_{\odot}}\right) \left({R \over 10^{10} {\rm cm}}\right)^{-1}
\left({H \over R}\right)^{2} \ {\rm K},
\end{equation}
which is clearly contradicted by observations.
 
When one calculates self-consistent (in the sense that $H_{\rm irr}$ in Eq. (13)
is calculated and not {\sl assumed}) models of irradiated discs one sees that the
outer disc regions whose structure could be affected by irradiation are hidden in the shadow
of the convex disc. This `self--screening' results from the same physical process that is
at the origin of the dwarf-nova instability: a dramatic change of opacities due
to hydrogen recombination. Therefore, although irradiation could stabilize the disc, the
unstable disc regions (see Fig. 3), cannot be irradiated by a point source located at the
disc mid-plane. Models invoking the stabilizing effects of irradiation (van Paradijs
1996; King \& Ritter 1998) have therefore to be revised (DLHC). Since outer disc
regions in low mass X-ray binaries are clearly irradiated, 
(van Paradijs \& McClintock 1995)
model revisions must concern the disc--irradiating source geometry.
To represent a geometry allowing the disc to see the irradiating source
DLHC assumed that 
\begin{equation}
T^4_{\rm irr} = {\cal C} {\dot M c^2 \over 4 \pi \sigma R^2}
\label{tirr2}
\end{equation} 
and calculated irradiated disc structure. The results are shown in Fig. 4. The continous
line represents both a non-irradiated disc and a disc irradiated according to Eq. (13)
with a self-consistently calculated $H_{\rm irr}$: there is no difference between the
two cases.

\section{Can irradiation by a hot white dwarf explain the UV--delay in dwarf novae?}

As shown by Hameury, Lasota \& Dubus (1999) the answer to this question asked by
King (1997) is `no'. Irradiation of the disc by the hot white dwarf may,
however, be important in a different context (see Sect. 6).

\section{Can the secondary in a dwarf-nova system be irradiated during outbursts ?}

The answer is `yes' at least for SS Cyg (Hessman 1984) and WZ Sge (Smak 1993). In
these systems it was {\sl observed} that the secondary's hemisphere facing the accreting
white dwarf was, during the outburst, heated to 16 000 - 17000 K, which for WZ Sge
implies that the irradiating flux is $\sim 5000$ times larger that the star's intrinsic
flux. It is hard to believe that the companion does not increase its mass transfer rate
in such conditions. WZ Sge is a very special system anyway (Lasota, Kuulkers \& Charles 1999).

\section{Can irradiation of the disc and of the secondary determine outburst properties?}

Warner (1998) suggested that outburst properties of
SU UMa stars could be explained by the effects of irradiation of both the accretion disc
and the secondary. Preliminary results by Hameury \& Lasota (1999, in preparation; see
also Hameury these proceedings) seem to confirm that properties of SU UMa (and ER UMa)
stars may be explained in this way.

\section*{Acknowledgement}

This article was written during a visit at the Weizmann Institute. I thank Moti Milgrom and 
the Department of Condensed Matter for hospitality and the Einstein Center for support.

\section{References}

\vspace{1pc}

\re
1.\ de Jong J.A., van Paradijs J., Augusteijn T. \ 1996, A\&A, 314, 484
\re
2.\ Dubus, G., Lasota, J.-P., Hameury, J.-M., Charles, Ph. \ 1999, MNRAS, in \ press
\re
3.\ Hameury J.-M., Lasota  J.-P., Hur\'e J.-M.\ 1997, MNRAS, 287, 937
\re
4.\ Hameury J.-M., Lasota  J.-P., Dubus G.\ 1999, MNRAS, in press
\re
5.\ Hameury J.-M., Menou K., Dubus G., Lasota J.-P., Hur\'e J.-M.\ 1998, MNRAS, 298, 1048
\re
6.\ Hessman F.V., Robinson E.L., Nather R.E., Zhang E.-H.\ 1984, ApJ, 286, 747
\re
7.\ Idan I., Shaviv G.\ 1996, MNRAS, 281, 604
\re 
8.\ King A.R.\ 1997, MNRAS, 288, L16
\re
9.\ King A.R., Ritter H. \ 1998, MNRAS, 293, 42
\re
10.\ Lasota J.-P., Kuulkers, E., Charles, Ph. \ 1998, MNRAS, submitted
\re
11.\ van Paradijs J.\ 1996, ApJ, 464, L139
\re
12.\ van Paradijs J., McClintock J.E.\ 1995, in  X-ray Binaries,
   eds.\ Lewin W.H.G., van Paradijs J., van den Heuvel E.P.J., 
   (Cambridge University Press, Cambridge) p. 58
\re
13.\ Shakura N.I., Sunyaev R.A.\ 1973, A\&A, 24, 337
\re
14.\ Shaviv G., Wehrse R.\ 1986, A\&A, 159, L5
\re
15.\ Shaviv G., Wehrse R.\ 1991, in Theory of Accretion Disks,
 eds.\ Meyer F., Duschl W.J., Frank J., Meyer-Hofmeister E.,
 (Kluwer, Dordrecht) p. 419
16.\ Smak J.\ 1993, Acta. Astron., 43, 101
\re
17.\ Tuchman Y., Mineshige S., Wheeler J.C.\ 1990, ApJ, 359, 164
\re
18.\ Vrtilek S.D. \ et al.\  1990, A\&A, 235, 165
\re
19.\ Warner B. \ 1998,  in Wild Stars in the Old West: Proceedings of the 13th
North American Workshop on Cataclysmic Variables and Related Objects, eds. \
S. Howell, E. Kuulkers \& C. Woodward, (ASP Conf. Ser. 137) p. 2

\end{document}